\documentclass[preprint,prd,superscriptaddress]{revtex4}
\usepackage{graphicx}% Include figure files
\usepackage{dcolumn}% Align table columns on decimal point
\usepackage{bm}% bold math
\newcommand{\p}{\partial}
\newcommand{\pslash}{p\kern-1ex /}
\newcommand{\lslash}{l\kern-1ex /}
\newcommand{\kslash}{k\kern-1ex /}
\newcommand{\dslash}{\p\kern-1.2ex /}
\newcommand{\Dslash}{{\cal D}\kern-1.5ex /}
\newcommand{\Aslash}{A\kern-1.2ex /}

\newcommand{\tr}{{\rm tr}}
\newcommand{\Tr}{{\rm Tr}}

\newcommand{\bea}{\begin{eqnarray}}
\newcommand{\eea}{\end{eqnarray}}
\newcommand{\vol}{\Omega}

\newcommand{\BAN}{\begin{eqnarray*}}
\newcommand{\EAN}{\end{eqnarray*}}
\begin{document}

\newcommand{\NTU}{
  Physics Department, 
  National Taiwan University, Taipei~10617, Taiwan  
}

\newcommand{\CQSE}{
  Center for Quantum Science and Engineering,  
  National Taiwan University, Taipei~10617, Taiwan  
}

\newcommand{\RCAS}{
  Research Center for Applied Sciences, Academia Sinica,
  Taipei~115, Taiwan
}

\preprint{NTUTH-11-505D}
 
\title{Topological Susceptibility in Two Flavors Lattice QCD 
       with the Optimal Domain-Wall Fermion}

\author{Ting-Wai~Chiu}
\affiliation{\NTU}
\affiliation{\CQSE}

\author{Tung-Han~Hsieh}
\affiliation{\RCAS}

\author{Yao-Yuan~Mao}
\affiliation{\NTU}

\collaboration{for the TWQCD Collaboration}
\noaffiliation


\begin{abstract}

  We determine the topological susceptibility of  
  the gauge configurations generated by lattice simulations 
  using two flavors of optimal domain-wall fermion on the $ 16^3 \times 32 $ 
  lattice with length 16 in the fifth dimension,    
  at the lattice spacing $ a \simeq 0.1 $~fm. 
  Using the adaptive thick-restart Lanczos algorithm, 
  we project the low-lying eigenmodes of the overlap 
  Dirac operator, and obtain the topological charge 
  of each configuration, for eight ensembles with 
  pion masses in the range $ 220-550 $ MeV. 
  From the topological charge, we compute the topological susceptibility  
%  $ \chi_t = \langle Q_t^2 \rangle /\vol $,  
  and the second normalized cumulant.
%  $ c_4 = -(\langle Q_t^4 \rangle - 3 \langle Q_t^2 \rangle^2 )/\vol $, 
%  where $ \vol $ is the volume of the lattice.
  Our result of the topological susceptibility 
  agrees with the sea-quark mass dependence predicted by the chiral perturbation theory 
  and provides a determination of the chiral condensate, 
  $\Sigma^{\overline{\mathrm{MS}}}(\mathrm{2~GeV})=[\mathrm{259(6)(7)~MeV}]^3 $, 
  and the pion decay constant $ F_\pi = 92(12)(2) $~MeV. 

\end{abstract}

\maketitle

\section{Introduction}

The vacuum of Quantum Chromodynamics (QCD) has a non-trivial topological structure.
The cluster property and the gauge invariance require that the ground state
must be the $\theta$ vacuum, a superposition of gauge configurations in
different topological sectors. The topological susceptibility
($ \chi_t $) is the most crucial quantity to measure the
topological fluctuations of the QCD vacuum,
which plays an important role in breaking the $ U_A(1) $ symmetry.
Theoretically, $ \chi_t $ is defined as
\bea
\label{eq:chi_t}
\chi_{t} = \int d^4 x  \left< \rho(x) \rho(0) \right>, 
\eea
where 
%\bea
%\label{eq:rho}
$ \rho(x) = 
\epsilon_{\mu\nu\lambda\sigma} \tr[ F_{\mu\nu}(x) F_{\lambda\sigma}(x) ]/(32 \pi^2) $, 
%\eea
is the topological charge density 
%($\propto$ axial anomaly)
expressed in term of the matrix-valued field tensor $ F_{\mu\nu} $.
With mild assumptions, Witten \cite{Witten:1979vv} and
Veneziano \cite{Veneziano:1979ec}
obtained a relationship between the topological susceptibility
in the quenched approximation and the mass of $ \eta' $ 
meson (flavor singlet) in the full QCD. 
%namely, 
%\BAN
%\chi_t(\mathrm{quenched}) = \frac{F_\pi^2 }{6}\left(M_{\eta'}^2 - 2 M_K^2 + M_\eta^2 \right),  \quad (N_f=3), 
%\EAN
%where $ F_\pi \simeq 93 $ MeV, the decay constant of pion.
This implies that the mass of $ \eta' $ is essentially due to
the axial anomaly relating to non-trivial topological
fluctuations, 
unlike those of the (non-singlet) approximate Goldstone bosons.

From (\ref{eq:chi_t}), we obtain
\bea
\label{eq:chit_Qt}
\chi_t = \frac{\left< Q_t^2 \right>}{\Omega}, \hspace{4mm}
Q_t \equiv  \int d^4 x \rho(x),
\eea
where $ \Omega $ is the volume of the system, and
$ Q_t $ is the topological charge (which is an integer for QCD).
Thus, one can determine $ \chi_t $ by counting the number of
gauge configurations for each topological sector.
Furthermore, we can also obtain the second normalized cumulant
\bea
\label{eq:c4}
c_4 = -\frac{1}{\vol} \left[   \langle Q_t^4 \rangle
                            -3 \langle Q_t^2 \rangle^2 \right],
\eea
which is related to the leading anomalous contribution to
the $ \eta'-\eta' $ scattering amplitude in QCD, as well as the
dependence of the vacuum energy on the vacuum angle $ \theta $.
(For a recent review, see, for example, Ref. \cite{Vicari:2008jw}
and references therein.)

However, for lattice QCD, it is difficult to extract $ \rho(x) $
and $ Q_t $ unambiguously from the gauge link variables, due to
their rather strong fluctuations.
To circumvent this difficulty, we may consider
the Atiyah-Singer index theorem
%\cite{Atiyah:1968mp,Atiyah:1984tf}
\cite{Atiyah:1968mp}, 
%\BAN
%\label{eq:AS_thm}
$ Q_t = n_+ - n_- = \mathrm{index}({\cal D}) $, 
%\EAN
where $ n_\pm $ is the number of zero modes of the massless Dirac
operator $ {\cal D} \equiv \gamma_\mu ( \partial_\mu + i g A_\mu) $
with $ \pm $ chirality.

For lattice QCD with exact chiral symmetry, it is well-known that
the overlap Dirac operator \cite{Neuberger:1997fp,Narayanan:1995gw}
in a topologically non-trivial gauge background
possesses exact zero modes (with definite chirality) satisfying
the Atiyah-Singer index theorem. Thus we can obtain the topological
charge from the index of the overlap Dirac operator.
Writing the overlap Dirac operator as
\BAN
\label{eq:overlap}
D_{o} = m_0 \left( 1 + \gamma_5 \frac{H_w}{\sqrt{H_w^2}} \right),
\EAN
where $ H_w $ is the standard Hermitian Wilson operator with negative
mass $ -m_0 $ ($ 0 < m_0 < 2 $), then its index is
\BAN
\label{eq:index_overlap}
 \mbox{index}(D)
= \Tr \left[ \gamma_5 \left( 1 - \frac{D_{o}}{2m_0} \right) \right]
%= -\frac{1}{2} \Tr \left( \frac{H_w}{\sqrt{H_w^2}} \right) 
= n_+ - n_-
= Q_t,
\EAN
where $ \Tr $ denotes trace over Dirac, color, and site indices. 

In this paper, we measure the topological charge 
of the gauge configurations generated by lattice simulations
of two flavors QCD on a $ 16^3 \times 32 $ lattice, 
with the optimal domain-wall fermion (ODWF) \cite{Chiu:2002ir} at $ N_s = 16 $, 
and plaquette gauge action at $ \beta = 5.95 $, for eight sea-quark masses
$ m_q a = 0.01, \cdots, 0.08 $ with the interval 0.01.   

Mathematically, ODWF is a theoretical framework which can preserve 
the chiral symmetry optimally for any given $N_s$, 
with a set of analytical weights $ \{ \omega_s, s = 1, \cdots, N_s \} $, 
one for each layer in the fifth dimension \cite{Chiu:2002ir}. 
Thus the artifacts due to the chiral
symmetry breaking with finite $ N_s $ can be reduced to the minimum, 
especially in the chiral regime.
The 4-dimensional effective Dirac operator of massless ODWF is
%\begin{widetext}
\bea
\label{eq:odwf_4d}
%\begin{aligned}
D = m_0 [1+ \gamma_5 S_{opt}(H_w) ], \quad
S_{opt}(H_w) = \frac{1-\prod_{s=1}^{N_s} T_s}{1 + \prod_{s=1}^{N_s} T_s}, \quad
T_s = \frac{1-\omega_s H_w}{1+\omega_s H_w},  
%\end{aligned}
\eea
%\end{widetext}
which is exactly equal to the Zolotarev optimal rational approximation 
of the overlap Dirac operator. That is,   
$ S_{opt}(H_w) = H_w R_Z(H_w) $, where $ R_Z(H_w)$ is the optimal 
rational approximation of $ (H_w^2)^{-1/2} $ 
\cite{Akhiezer:1992, Chiu:2002eh}.

We use the adaptive thick-restart Lanczos algorithm \cite{a-TRLan}
to project the low-lying eigenmodes of the 4-dimensional
effective Dirac operator (\ref{eq:odwf_4d}), 
and obtain the topological charge $ Q_t $
of each gauge configuration.
Then we compute the topological susceptibilty
$ \chi_t $ and the second normalized cumulant $ c_4 $, 
and compare our results to the Chiral Perturbation Theory (ChPT).
We summarize the ChPT formulas as follows.  

In 1992, Leutwyler and Smilga \cite{Leutwyler:1992yt} derived
the relationship between $ \chi_t $ and the quark mass,
at the leading order in ChPT. For 2 flavors QCD, it reads
\bea
\label{eq:chit_ChPT_tree}
\chi_t =  \Sigma \left( m_u^{-1} + m_d^{-1} \right)^{-1}, 
\eea
where $ m_u$, $m_d$ are the quark masses,
and $ \Sigma $ is the chiral condensate.
This implies that in the chiral limit ($ m_u \to 0 $)
the topological susceptibility is suppressed by the internal quark loops.
Most importantly, (\ref{eq:chit_ChPT_tree}) provides a viable way
to extract $ \Sigma $ from $ \chi_t $ in the chiral regime. 

Recently, the topological susceptibility has been derived to the one-loop order
in ChPT for an arbitrary number of flavors \cite{Mao:2009sy}.
For $ N_f = 2 $ with degenerate $u$ and $d$ quark masses
($m_u = m_d \equiv m_q $), the foumula reduces to
%\begin{widetext}
\bea
\label{eq:chitomq_ChPT_NLO_nf2}
%\begin{aligned}
\frac{\chi_t}{m_q} = \frac{\Sigma}{2} \bigg\{ 1 
- 3 \left(\frac{\Sigma m_q}{16 \pi^2 F_\pi^4}\right) \ln \left( \frac{2 \Sigma m_q}{ F_\pi^2 \mu_{sub}^2} 
\right) + 32 \left(\frac{\Sigma}{F_\pi^4} \right) (2 L_6 + 2 L_7 + L_8) m_q \bigg\},
%\end{aligned}
\eea
%\end{widetext}
where $ L_i $ are renormalized low-energy coupling constants defined at $ \mu_{sub} $ \cite{Gasser:1984gg}.
In this paper, we fix $ \mu_{sub} = 770 $ MeV.

Furthermore, the second normalized cumulant $ c_4 $ has been 
derived in ChPT at the tree-level for an arbitrary number of flavors 
\cite{Mao:2009sy,Aoki:2009mx}. 
For 2 flavors QCD, it reads
\bea
\label{eq:c4_ChPT_tree}
c_4 = - \Sigma \left( m_u^{-3} + m_d^{-3} \right)
\left( m_u^{-1} + m_d^{-1} \right)^{-4}.    
\eea
In the isospin limit ($ m_u = m_d $), the ratio $ c_4 / \chi_t $ goes to $ -1/4 $. 

In this paper, we investigate to what extent the quark mass dependences of
$ \chi_t $ in lattice QCD with optimal domain-wall fermion would agree with
the ChPT to the one-loop order, and to determine $ \Sigma $ and $ F_\pi $ 
from our data of $ \chi_t $.
In principle, we can also extract $ \Sigma $ from $ c_4 $, however, this would require
much higher statistics than that of $ \chi_t $.

\section{Lattice Setup}

Simulations are carried out for two flavors QCD on a $16^3 \times 32$
lattice at the lattice spacing $a \sim 0.1$~fm, for eight sea-quark masses 
$ m_q a =0.01, 0.02, 0.03, 0.04, 0.05, 0.06, 0.07$, and $ 0.08 $ respectively.
For the quark part, we use the optimal domain-wall fermion with $ N_s = 16 $.
For the gluon part, we use the plaquette action at 
$\beta = 5.90$ and $ \beta=5.95 $ respectively. 
An outline of our simulation algorithm and its acceleration with Nvidia GPUs 
has been presented in Refs. \cite{Chiu:2009wh,Chiu:2011rc}, 
and the details will be presented in Ref. \cite{Chiu:2011si}.
Our preliminary physical results of the $ \beta = 5.90 $ ensemble 
have been presented in Refs. \cite{Chen:2011qy,Hsieh:2011qx}.
In this paper, we present our results of the topological susceptibility 
$ \chi_t $ and the second normalized cumulant $ c_4 $ 
of the $ \beta=5.95 $ ensemble. 
 
For each sea-quark mass, we perform hybrid Monte-Carlo simulations on 
30 GPUs independently, with each GPU generating 400 trajectories.   
After discarding 300 trajectories for thermalization, each GPU yields 
100 trajectories. Thus, with 30 GPUs running independently,  
we accumulated total $ 3000 $ trajectories for each sea-quark mass. 
From the saturation of the binning error of the plaquette, as well as
the evolution of the topological charge, we estimate the autocorrelation time
to be around 10 trajectories. Thus we sample one configuration every
10 trajectories, then we have $ 300 $ configurations for each sea-quark mass.
With a GPU cluster of 250 GPUs, we can simulate 8 sea-quark masses concurrently. 
It takes about 5 months to complete the simulations for the $ \beta = 5.95 $ ensemble.

We determine the lattice spacing by heavy quark potential 
with Sommer parameter $ r_0 = 0.49 $~fm.
%The lattice spacing versus the quark mass is plotted in Fig.~\ref{fig:a_mq_b595}.
Using the linear fit, we obtain the lattice spacing in the chiral limit,
$ a = 0.1032(2) $~fm, which gives $ a^{-1} = 1.911(4)(6) $~GeV, where 
the systematic error is estimated with the uncertainty of $ r_0 $.

%\begin{figure}[htb]
%\begin{center}
%\centerline{\includegraphics*[height=5.8cm,width=7.55cm]{a_mq_b595.eps}}
%\centerline{\includegraphics[width=90mm,clip=true]{a_mq_b595.eps}}
%\end{center}
%\caption{
%\label{fig:a_mq_b595}
%The lattice spacing $ a $~fm versus $ m_q a $ for two flavors QCD with ODWF.}
%\end{figure}

For each configuration, we calculate the zero modes plus 80 conjugate pairs of 
the lowest-lying eignmodes of the overlap Dirac operator. 
We outline our procedures as follows.
First, we project 240 low-lying eigenmodes of $ H_w^2 $ using adaptive 
thick-restart Lanczos alogorithm ($a$-TRLan) \cite{a-TRLan}, 
where each eigenmode has a residual less than $ 10^{-12} $.
Then we approximate the sign function of the overlap operator
by the Zolotarev optimal rational approximation with 64 poles,
where the coefficents are fixed with $ \lambda_{max}^2 = (6.4)^2 $,
and $ \lambda_{min}^2 $ equal to the maximum of
the 240 projected eigenvalues of $ H_w^2 $.
Then the sign function error is less than $ 10^{-14} $.
Using the 240 low-modes of $ H_w^2 $ and the Zolotarev approximation
with 64 poles, we use the $a$-TRLan algorithm again to 
project the zero modes plus 80 conjugate pairs of
the lowest-lying eignmodes of the overlap operator, 
where each eigenmode has a residual less than $ 10^{-12} $.
We store all projected eigenmodes for the later use.
In this paper, we use the index of the zero modes
to compute $ \chi_t $ and $ c_4 $.

\begin{figure}[!htb]
\begin{center}
\begin{tabular}{@{}cccc@{}}
\includegraphics*[height=5cm,width=3.8cm,clip=true]{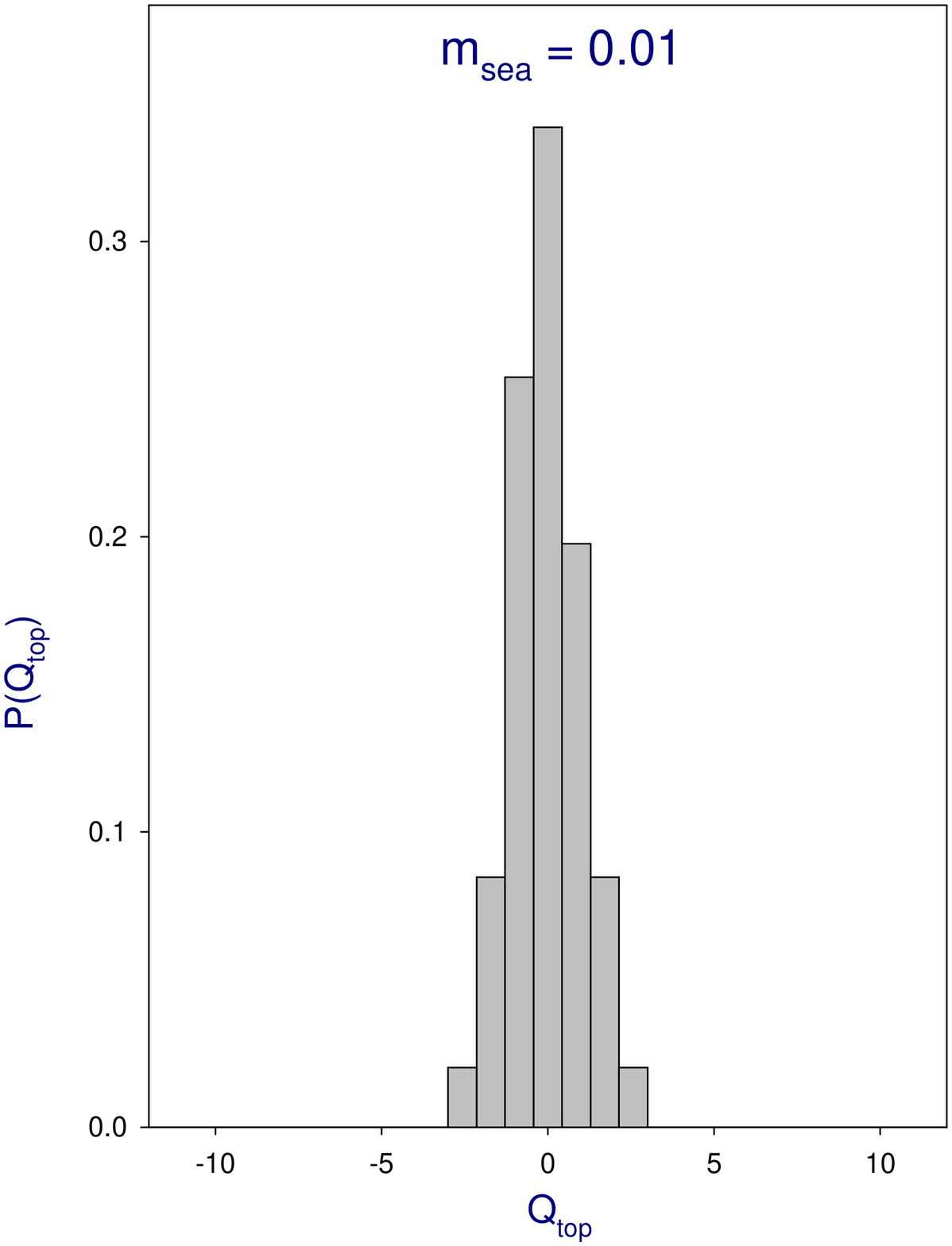}
&
\includegraphics*[height=5cm,width=3.8cm,clip=true]{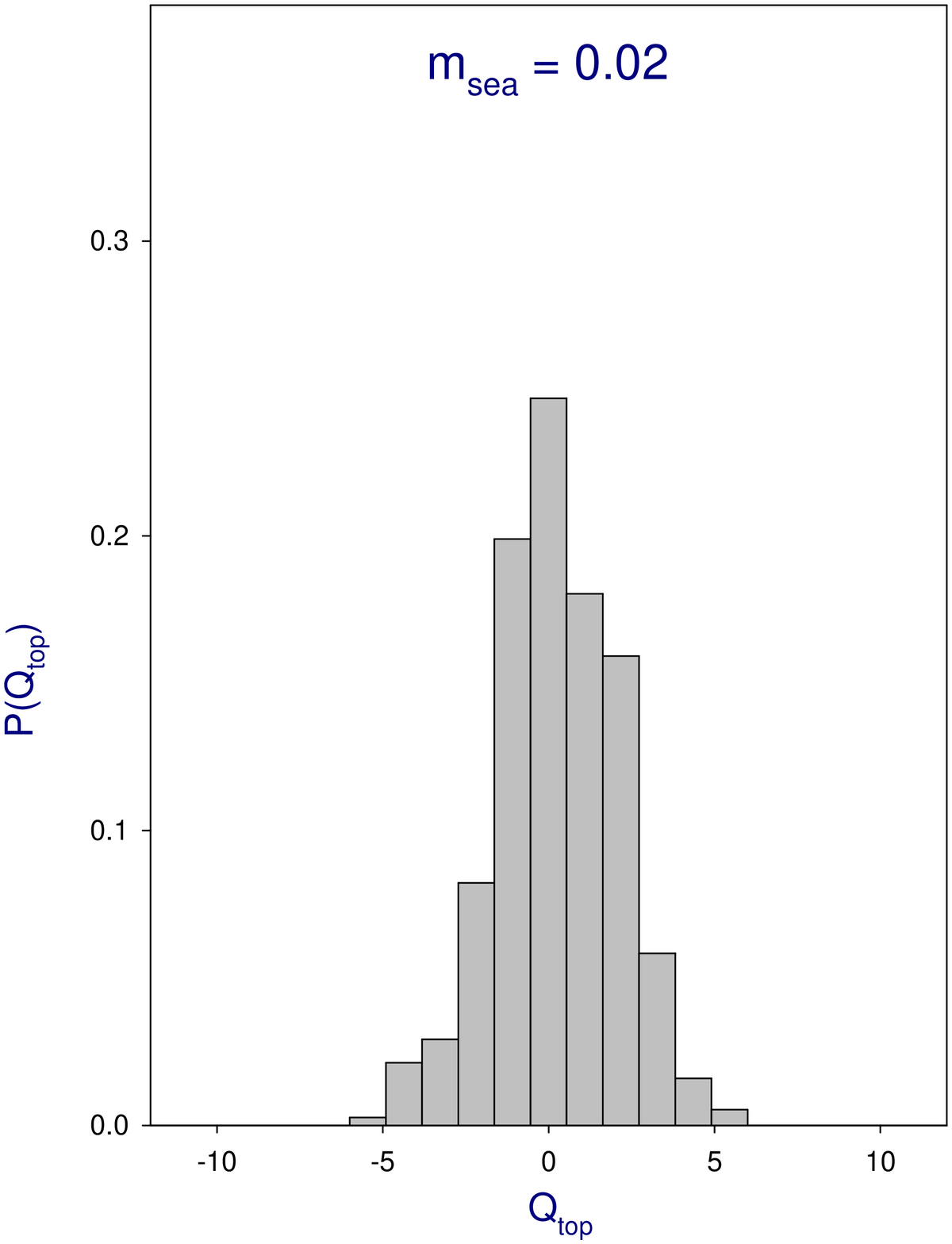}
&
\includegraphics*[height=5cm,width=3.8cm,clip=true]{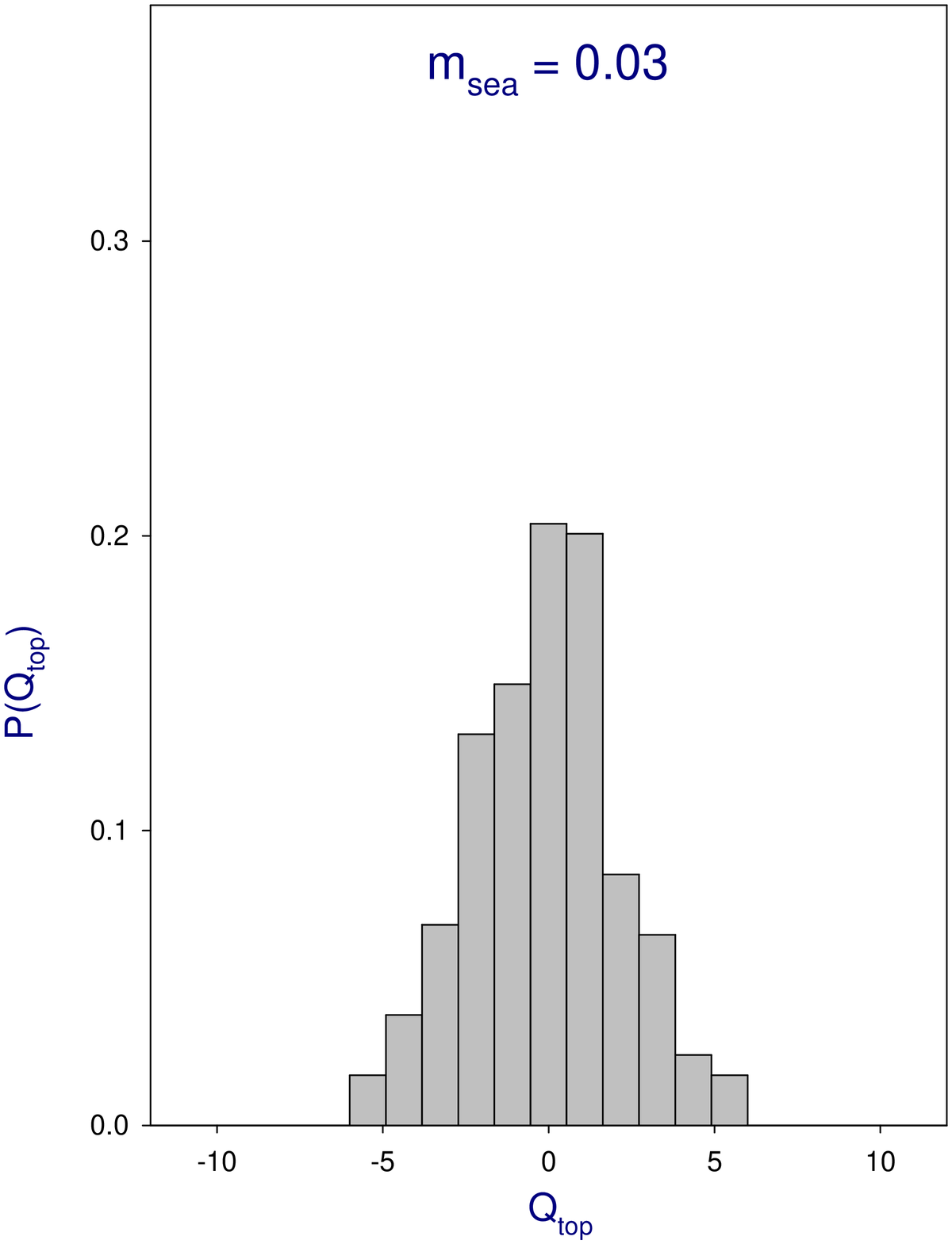}
&
\includegraphics*[height=5cm,width=3.8cm,clip=true]{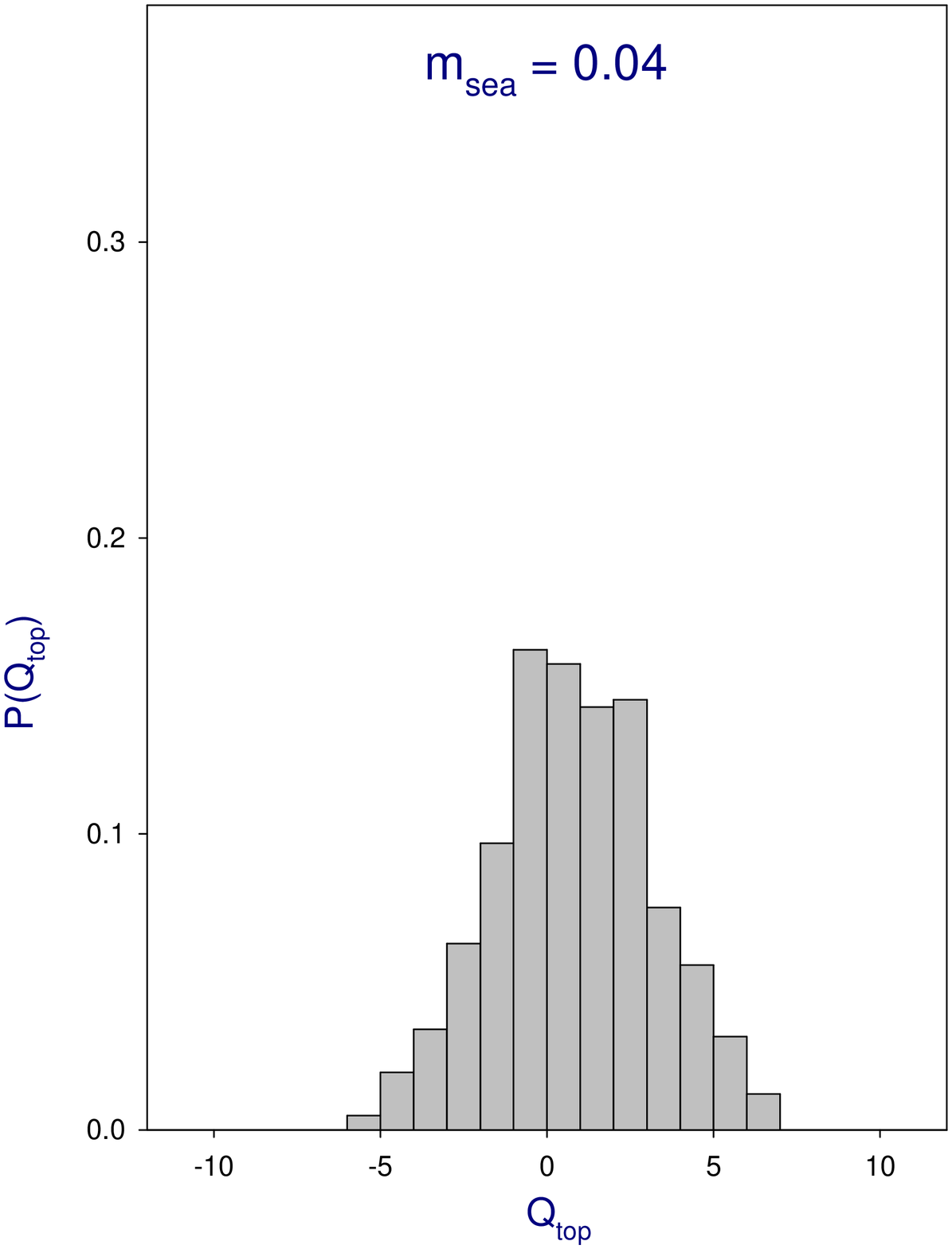}
\\
\includegraphics*[height=5cm,width=3.8cm,clip=true]{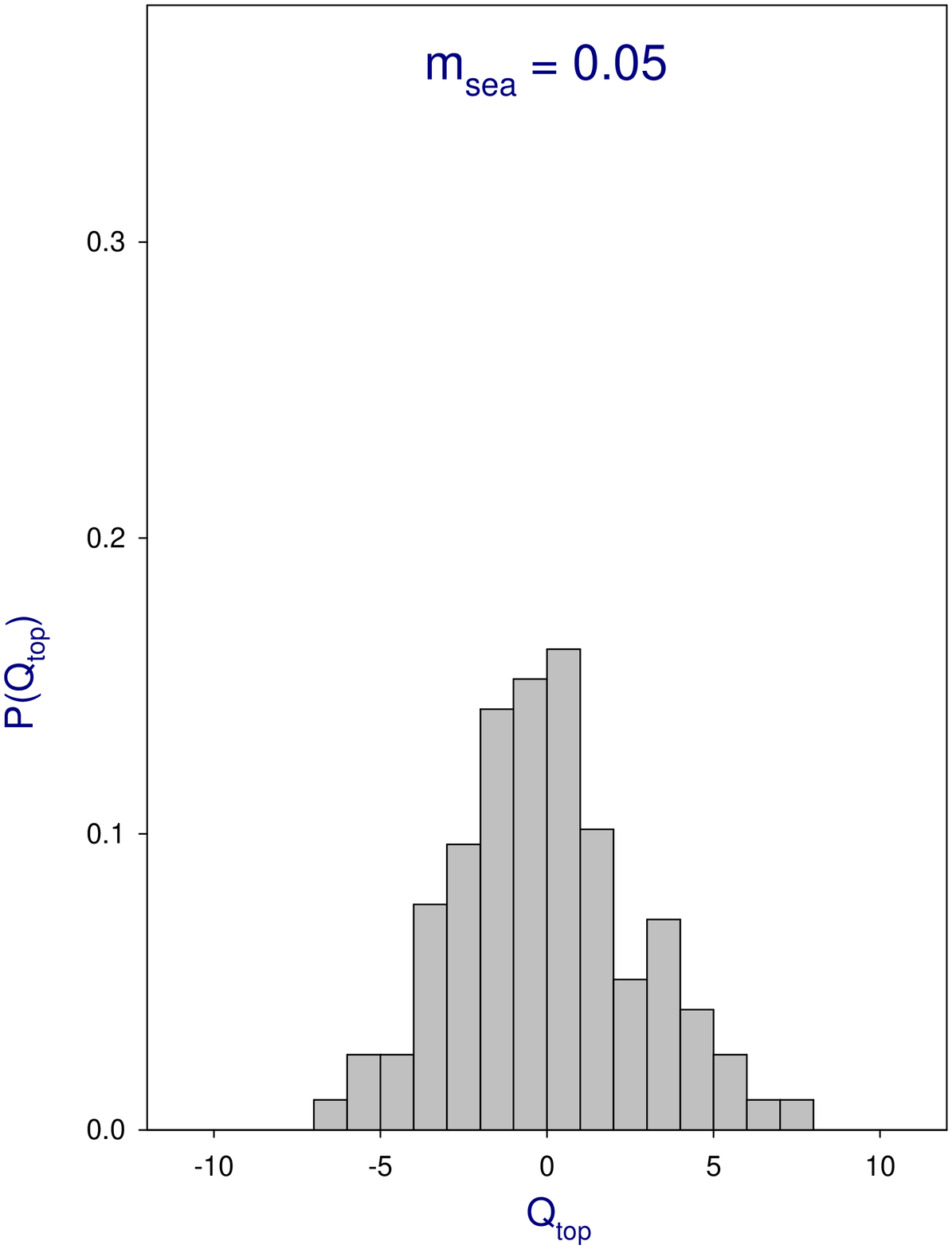}
&
\includegraphics*[height=5cm,width=3.8cm,clip=true]{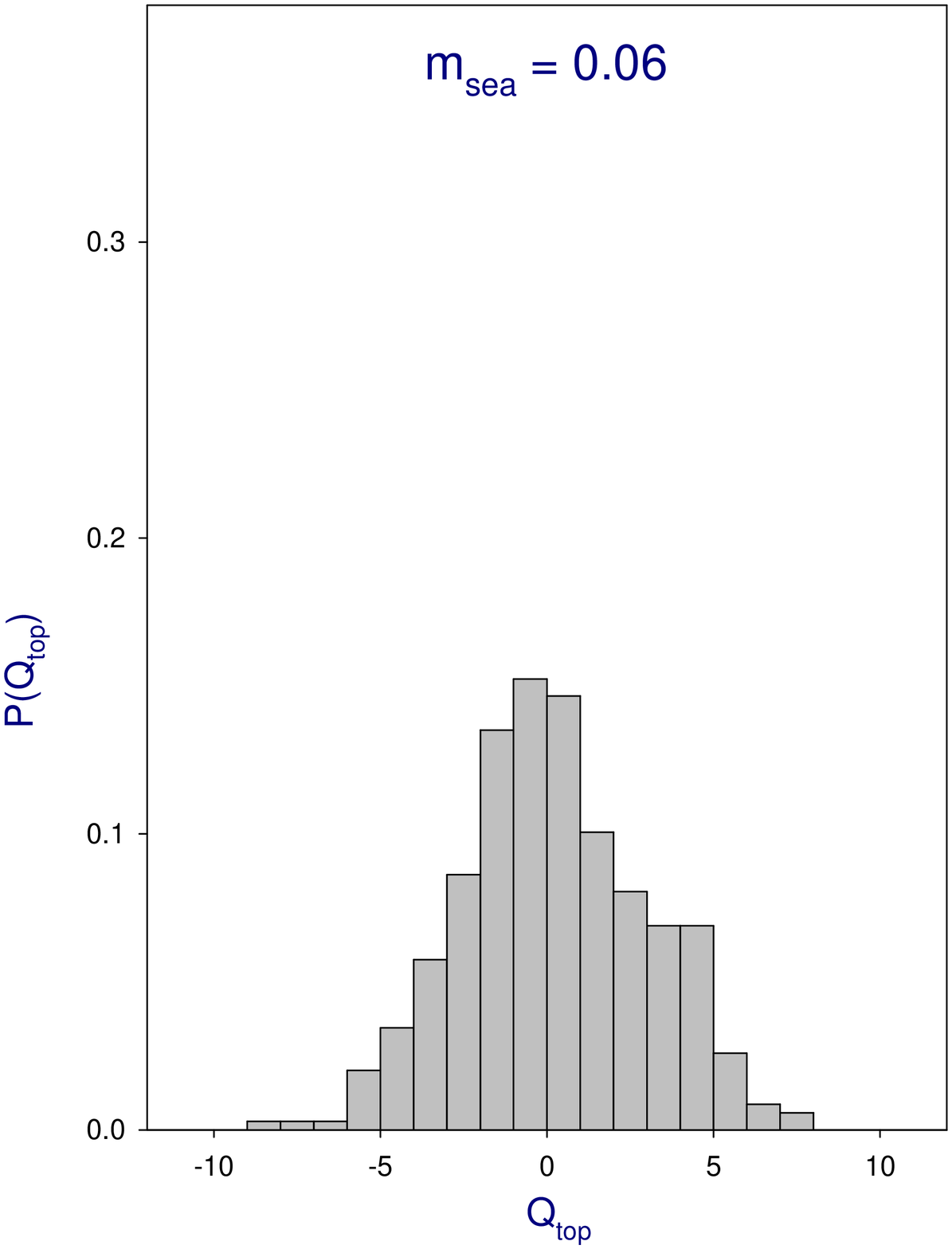}
&
\includegraphics*[height=5cm,width=3.8cm,clip=true]{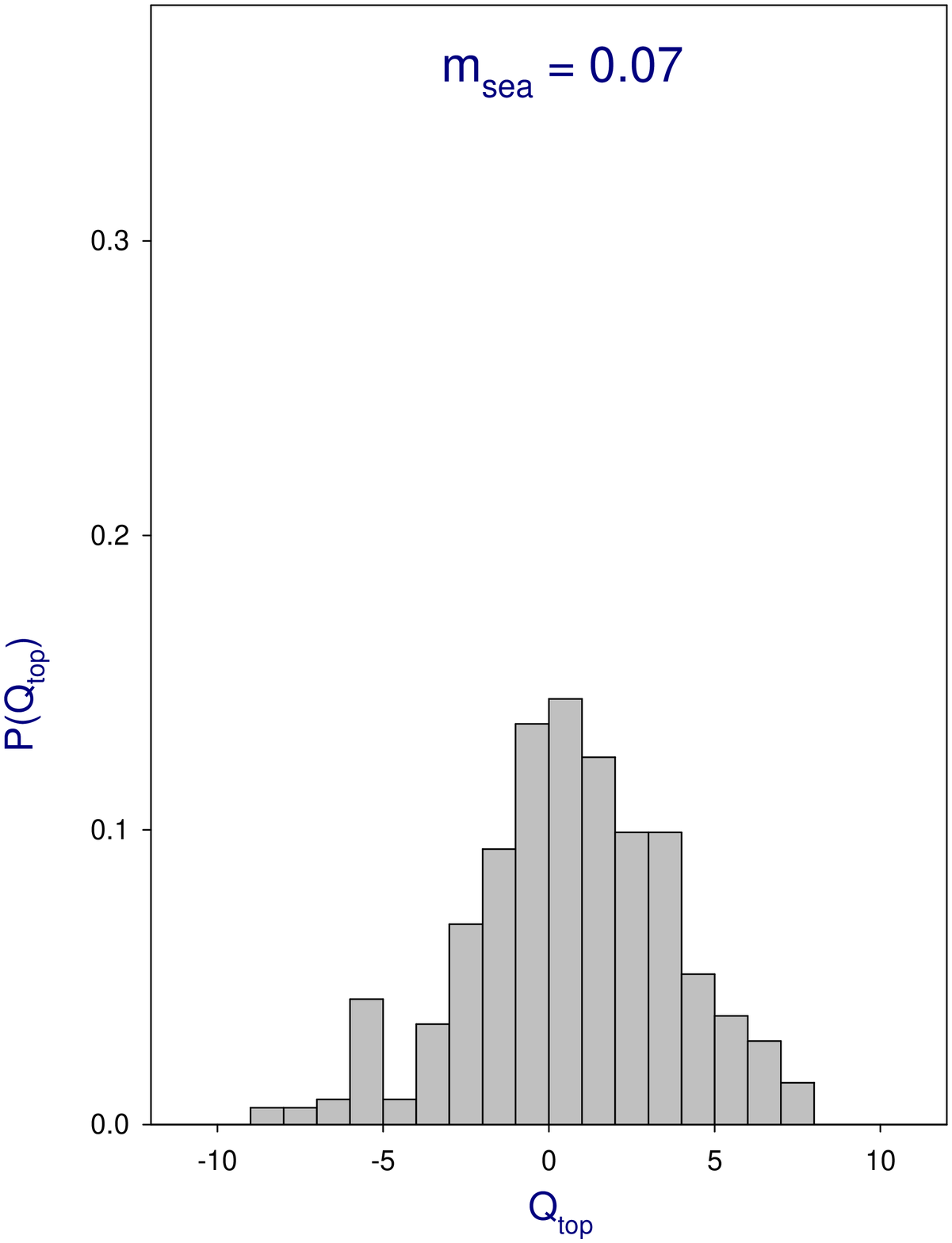}
&
\includegraphics*[height=5cm,width=3.8cm,clip=true]{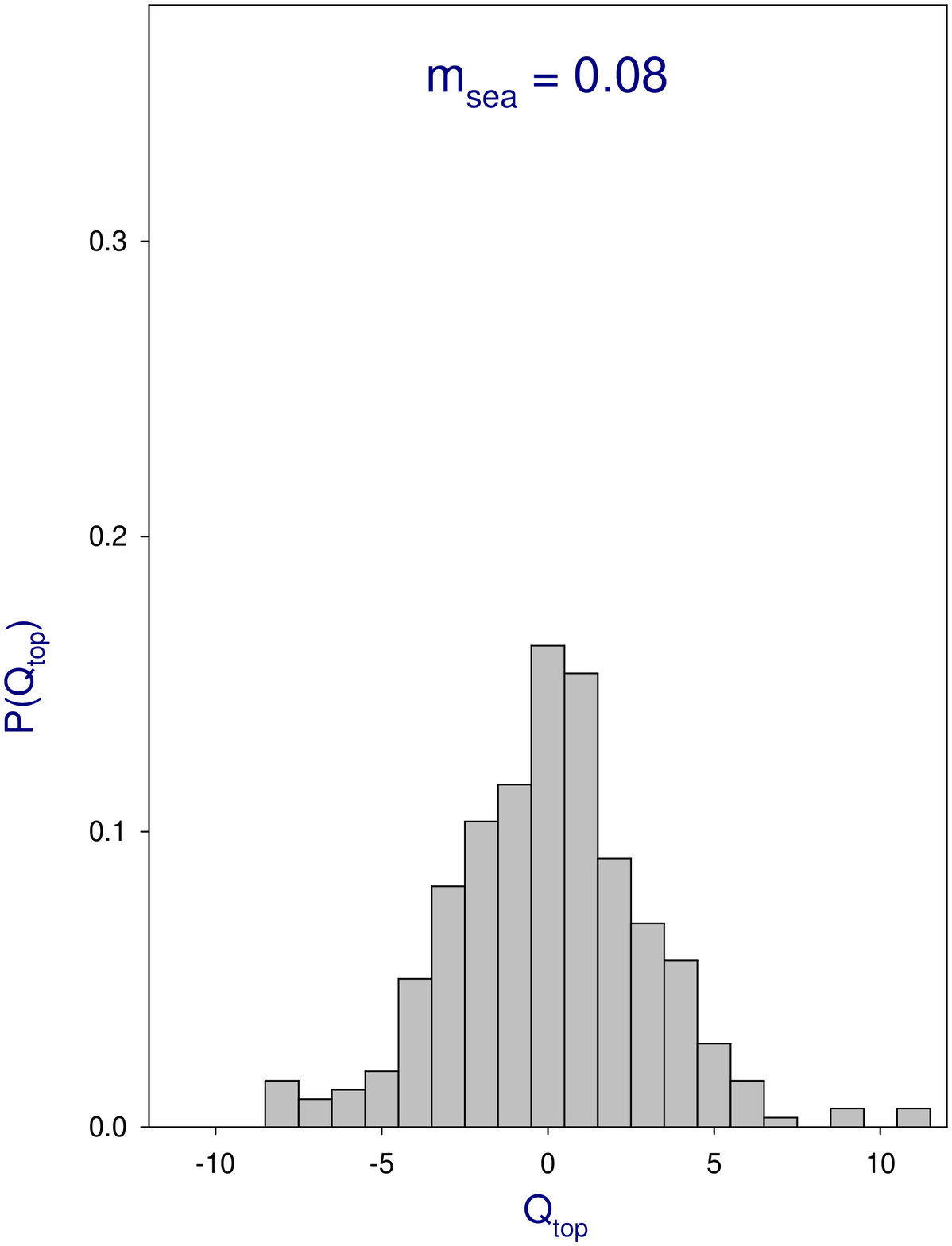}
\\
%\\ (a) & (b)
\end{tabular}
\caption{Histogram of topological charge distribution
for eight sea quark masses, $ m_q a =0.01, 0.02, 0.03, 0.04, 0.05, 0.06, 0.07$,
and $ 0.08 $ respectively.
}
\label{fig:Q_hist}
\end{center}
\end{figure}

\begin{table}[th]
\begin{center}
\begin{tabular}{|c|ccccc|}
\hline
$ m_q a $ & $ \chi_t $ & & $ c_4 $ & & $ c_4/\chi_t $  \\
\hline
0.01  & $ 1.13(10) \times 10^{-5} $ & & $ -1.22(1.15) \times 10^{-5} $ & & $ -1.07(1.01) $     \\
0.02  & $ 2.24(18) \times 10^{-5} $ & & $ -3.79(2.77) \times 10^{-5} $ & & $ -1.69(1.24) $     \\ 
0.03  & $ 3.29(27) \times 10^{-5} $ & & $ -1.25(2.87) \times 10^{-5} $ & & $ -0.04(0.87) $     \\
0.04  & $ 4.40(30) \times 10^{-5} $ & & $  6.39(4.16) \times 10^{-5} $ & & $  1.45(95) $     \\
0.05  & $ 5.31(41) \times 10^{-5} $ & & $  5.75(7.96) \times 10^{-5} $ & & $  1.08(1.50) $     \\
0.06  & $ 6.04(44) \times 10^{-5} $ & & $  1.00(1.09) \times 10^{-4} $ & & $  1.66(1.81) $     \\
0.07  & $ 7.24(55) \times 10^{-5} $ & & $ -5.12(134)  \times 10^{-6} $ & & $ -0.07(1.86) $     \\
0.08  & $ 7.01(79) \times 10^{-5} $ & & $ -7.10(6.58) \times 10^{-4} $ & & $ -10.13(9.46) $     \\
\hline
\end{tabular}
\end{center}
\caption{
The topological susceptibility $ \chi_t $,
the second normalized cumulant $ c_4 $, and their ratio $ c_4/\chi_t $, 
versus the sea quark masses, for
$ N_f = 2 $ lattice QCD with the optimal domain-wall fermion.
}
\label{tab:chit_c4_mq}
\end{table}

\begin{figure}[tb]
\centering
\includegraphics[width=10cm,height=7cm,clip=true]{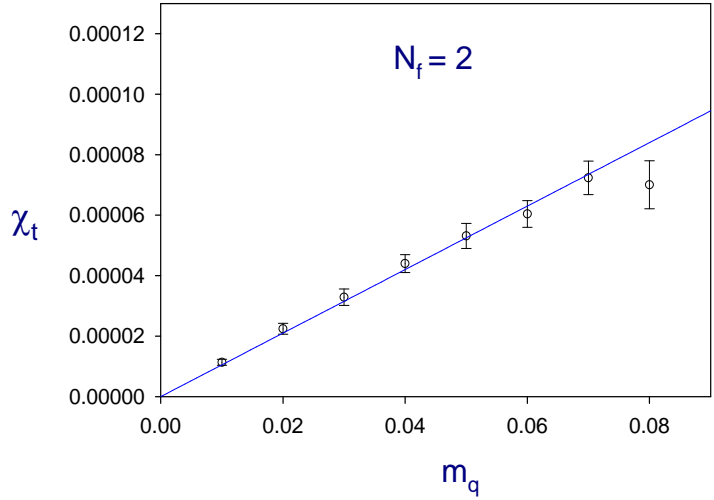}
\caption{The topological susceptibility $ \chi_t $
         versus the sea-quark mass $ m_q $ for 2 flavors lattice QCD 
         with ODWF. The straight line is the fit with the
         LO ChPT (\ref{eq:chit_ChPT_tree}).}
\label{fig:chit_mq_b595}
\end{figure}

\begin{figure}[tb]
\centering
\includegraphics[width=10cm,height=7cm,clip=true]{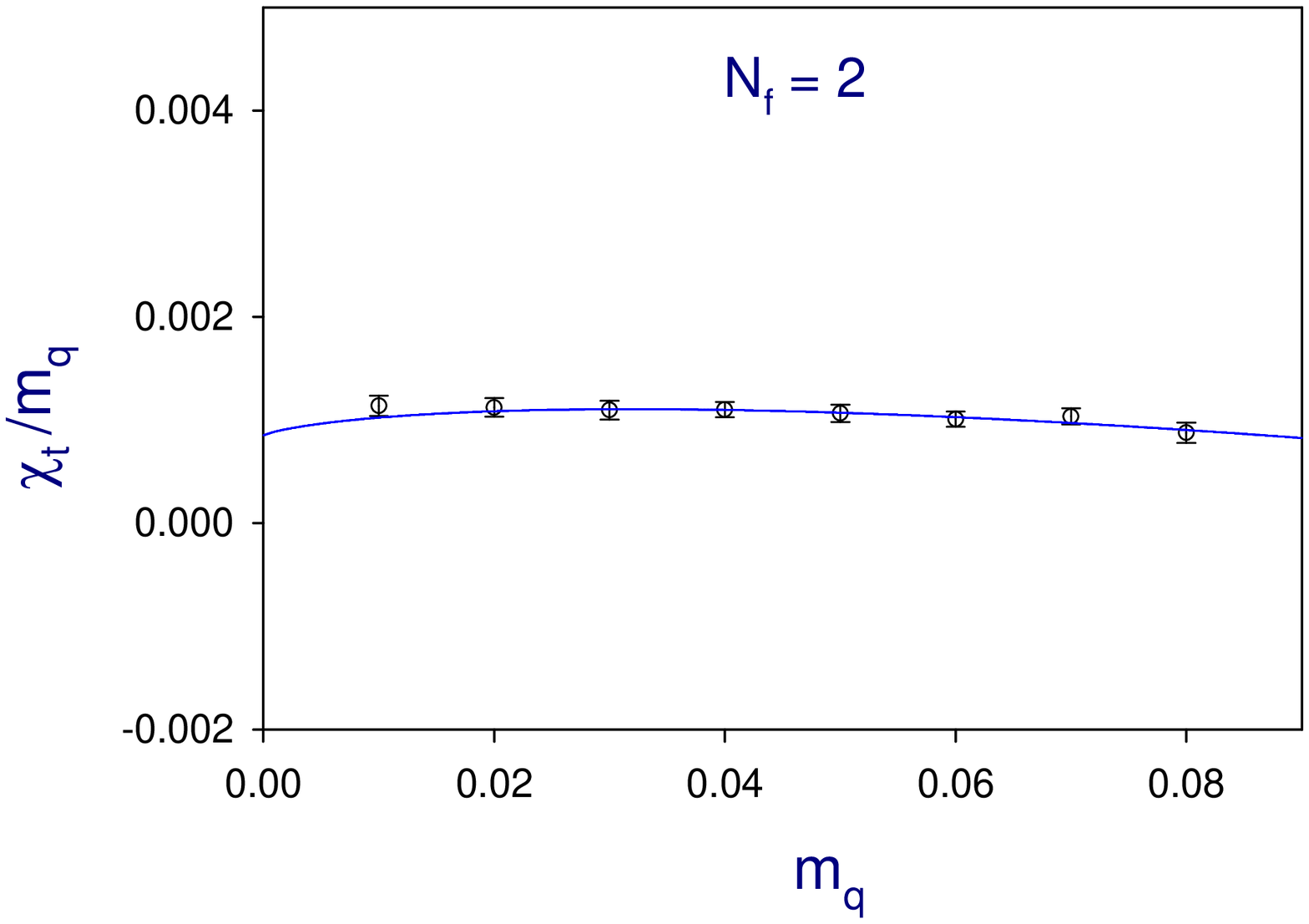}
\caption{The ratio $ \chi_t/m_q $ versus the sea-quark mass $ m_q $ 
         for 2 flavors lattice QCD with ODWF. 
         The solid line is the fit with the 
         NLO ChPT (\ref{eq:chitomq_ChPT_NLO_nf2}).}
\label{fig:chitomq_mq_b595}
\end{figure}

\section{Results}

In Fig.~\ref{fig:Q_hist}, we plot the histogram of topological charge    
distribution for $ m_q a =0.01, 0.02, \cdots, 0.08 $ respectively. 
Evidently, the probability distribution of $ Q_t $ for each sea-quark mass  
behaves like a Gaussian, and it becomes more sharply peaked around 
$ Q_t = 0 $ as the sea-quark mass $ m_q $ gets smaller. 

Using the result of $ Q_t $, we compute the topological susceptibility 
$\chi_t $ (\ref{eq:chit_Qt}), and the second normalized cumulant $ c_4 $ (\ref{eq:c4}).   
In Table \ref{tab:chit_c4_mq}, we list our results of $ \chi_t $, 
$ c_4 $, and the ratio $ c_4 / \chi_t $.  
The error is estimated using the jackknife method with bin size of
10 configurations, with which the statistical error saturates.

Evidently, the statistical error of the topological susceptibility 
is about 10\%, while that of $ c_4 $ is very large due to low statistics. 
Therefore, we cannot draw any conclusions from our result of $ c_4 $,  
as well as from the ratio $ c_4 / \chi_t $. 

In Fig \ref{fig:chit_mq_b595}, 
we plot our data of $ \chi_t $ versus the sea quark mass $ m_q $. 
The data points of $ \chi_t $ are well fitted by  
the Leutwyler and Smilga formula (\ref{eq:chit_ChPT_tree}) 
with $ \Sigma a^3 = 0.00200(15) $.
The fitted curve is plotted as the solid line in      
Fig \ref{fig:chit_mq_b595}.  

In Fig. \ref{fig:chitomq_mq_b595}, 
we plot our data of $ \chi_t/m_q $ versus the sea-quark mass $ m_q $. 
The data points of $ \chi_t/m_q $ are well fitted by  
the NLO ChPT formula (\ref{eq:chitomq_ChPT_NLO_nf2}) with
$ \Sigma a^3 = 0.0020(2) $, $ F_\pi a = 0.048(7) $, and 
\bea
\label{eq:L678}
(2 L_6 + 2 L_7 + L_8) = -0.0001(3), 
\eea   
where $ \mu_{sub} = 770 $~MeV  has been used. 
Using $ a^{-1} = 1.911(4)(6) $~GeV, we obtain 
$ \Sigma = [241(6)(1)~\mbox{MeV}]^3 $, and
\bea
\label{eq:Fpi_nf2}
F_\pi = 92(12)(2)~\mbox{MeV}, 
\eea  
where the errors represent a combined statistical error 
and the systematic error respectively. 

In order to convert $\Sigma$ to that in the
$\overline{\mathrm{MS}}$ scheme, we calculate the renormalization factor 
$Z_s^{\overline{\mathrm{MS}}}(\mathrm{2~GeV}) $ 
using the non-perturbative renormalization technique
through the RI/MOM scheme \cite{Martinelli:1994ty}, 
and our result is \cite{Chiu:2011np}
\BAN
Z_s^{\overline{\mathrm{MS}}}(\mathrm{2~GeV}) = 1.244(18)(39).  
\EAN
Then the value of $ \Sigma $ is transcribed to
\bea
\label{eq:Sigma_nf2}
\Sigma^{\overline{{\mathrm{MS}}}}(\mathrm{2~GeV})
  =[\mathrm{259(6)(7)~MeV}]^3,    
\eea
where the errors represent a combined statistical error
($a^{-1}$ and $Z_s^{\overline{\mathrm{MS}}}$) and
the systematic error respectively.
Since the present calculation is done at a single lattice spacing,
the discretization error cannot be quantified reliably, but
we do not expect much larger error because the optimal domain-wall fermion 
action is free from $O(a)$ discretization effects.
Our result of $ \Sigma $ (\ref{eq:Sigma_nf2}) is in good agreement with 
that extracted from $ \chi_t $ in (2+1) flavors QCD 
with domain-wall fermion \cite{Chiu:2008jq}, 
as well as with those extracted from $ \chi_t $ in $ N_f = 2 $ and $ N_f = 2+1 $ 
lattice QCD with overlap fermion in a fixed topology \cite{Aoki:2007pw, Chiu:2008kt}.

\section{Concluding remark}

To summarize, we measure the topological charge
of the gauge configurations generated by lattice simulations
of 2 flavors QCD with the optimal domain-wall fermion
at $ N_s = 16 $ and plaquette gauge action at $ \beta = 5.95 $, 
on a $ 16^3 \times 32 $ lattice.
We use the adaptive thick-restart Lanczos algorithm to compute 
the low-lying eigenmodes of the overlap Dirac operator, 
and obtain the topological charge of each gauge configuration, 
and from which we compute the topological susceptibility
for 8 sea-quark masses, each of $ 300 $ configurations. 
Our result of the topological susceptibility agrees with the 
sea-quark mass dependence predicted by
the NLO ChPT formula (\ref{eq:chitomq_ChPT_NLO_nf2}), 
and gives the first determination of both
the pion decay constant (\ref{eq:Fpi_nf2}) 
and the chiral condensate (\ref{eq:Sigma_nf2}) 
simultaneously from the topological susceptibility. 
Moreover, this study shows that it is feasible  
to perform large-scale simulations of unquenched lattice QCD, 
which not only preserve the chiral symmetry to a high precision,   
but also sample all topological sectors ergodically.

\begin{acknowledgments}
  This work is supported in part by the National Science Council
  (Nos.~NSC96-2112-M-002-020-MY3, NSC99-2112-M-002-012-MY3, NSC96-2112-M-001-017-MY3, 
  NSC99-2112-M-001-014-MY3, NSC99-2119-M-002-001) and NTU-CQSE~(Nos.~99R80869,~99R80873).
  We also thank NCHC and NTU-CC for providing facilities to perform part of our calculations.   
\end{acknowledgments}

\end{document}